\documentclass[12pt]{article}
\usepackage{latexsym}
\usepackage{amsmath}
\usepackage{amssymb}
\usepackage{amsthm}
\usepackage{dsfont}
\usepackage{longtable}
\usepackage{color}
\usepackage{verbatim}
\usepackage[sort&compress,numbers]{natbib}
\usepackage{graphicx}
\usepackage{epstopdf}
\usepackage{float}
\usepackage{wrapfig}

\begin{document}

\title{Three-parameter (two-sided) deformation of Heisenberg algebra}
\author{A.M. Gavrilik and I.I. Kachurik}
\maketitle

 \centerline{Bogolyubov Institute for Theoretical Physics of NAS of Ukraine}
 \centerline{14b, Metrologichna Str., Kyiv 03680, Ukraine}

\vspace{3mm}
\begin{abstract}
A 3-parametric two-sided deformation of Heisenberg algebra (HA),
with $p,q$-deformed commutator in the l.h.s. of basic defining
relation and certain deformation of its r.h.s., is introduced and
studied. The third deformation parameter $\mu$ appears in an extra
term in the r.h.s. as pre-factor of Hamiltonian.
 For this deformation of HA we find novel properties.
 Namely, we prove it is possible to realize this ($p,q,\mu$)-deformed
 HA by means of some deformed oscillator algebra.
    Also, we find the unusual property that the deforming factor $\mu$
 in the considered deformed HA {\em inevitably} depends explicitly
 on particle number operator $N$.
   Such a novel $N$-dependence is special for the two-sided
   deformation of HA treated jointly with its deformed oscillator realizations.
\end{abstract}

\vspace{3mm}

\vspace{2mm}

\newpage
\section{Introduction}

As mentioned in~\cite{Kempf}, W. Heisenberg was the first who
admitted consideration of generalized/modified versions of his
celebrated uncertainty relation which is a direct consequence of the
basic commutation rule
  \begin{equation}                  \label{HA}
 [X, P] = {\rm i}\hbar \
\end{equation}
 for the position and momentum operators in the Heisenberg algebra (HA).
  In this paper, we introduce rather general so-called two-sided deformation
  of HA. In it, not only the commutator in the l.h.s. of the
basic defining relation (DR) undergoes the procedure of deformation
 through replacing it with $q$-commutator as considered in    ref.~\cite{Ch-K}
 or $q,p$-commutator as studied here, but also its r.h.s. is
modified through introducing an extra term which involves
Hamiltonian multiplied with a pre-factor denoted by $\mu$.

   We study such nontrivially deformed system and show that it
 manifests some interesting properties.
 We first show how to realize the deformed HA through
 definite nonstandard deformed oscillator algebra.
 Next, as an unexpected thing, we find that
  the deformation factor (parameter) $\mu$ in front of Hamiltonian $H$
  in the r.h.s. of our basic DR {\em inevitably} becomes depending not only
  on the deformation parameter(s) $q$ (or $q,p$) of the $q$- (or $q,p$-)commutator
 in the l.h.s. of basic DR, but also on $N$ - the particle number operator.
 The disclosed property is characteristic just for this
 "two-sided"\ version of deformed Heisenberg algebra when it is
 realized through particular deformed oscillator algebra.
  It obviously cannot appear in cases involving the standard
  HA or its simple (one-parameter, one-sided) deformed versions.
%\vspace{2mm}

\section{Modified Heisenberg algebra involving the Hamiltonian}
%\medskip

The most natural  %simple-minded
extension of (\ref{HA}) may have the following form that involves a
function of the Hamiltonian ${\cal H}$:
\begin{equation}                    \label{f-HA}
[X, P] = {\rm i} \hbar f({\cal H})\ .
\end{equation}
Here $f({\cal H})$ is an operator function of ${\cal H}$ for
 which a series expansion is valid.
 With only constant term (zeroth order in ${\cal H}$),
 the relation (\ref{f-HA}) reduces to the usual HA.

Along these lines,
           Saavedra and Utreras in                      \cite{Saaved}
  (see also                                      \cite{Brod,Jan,Leiva,Bagchi}),
  with the goal of extension of quantum mechanics
  to the domain of high energy physics and in particular quark physics,
have considered a realization of this in the form linear in {\cal H}
($\mu $ is real):
\begin{equation}                              \label{mu-HA}
[X, P] = {\rm i} \hbar (1 + \mu {\cal H})\ .
\end{equation}
On the other hand,       Jannussis               in Ref.\,\cite{Jan}
 presented a rather
exotic deformation of the oscillator algebra - so-called
$\mu$-oscillator.
  For the $\mu$-deformed oscillator,
  the deformation structure function\footnote{See     Refs.\,\cite{Melj,Bona}
  for that concept;
     see also e.g.
     Refs.\,\cite{Bona,Chang,Man'ko,Plyush,AGI-1,AGI-2,Liu,AdGa,SIGMA,G-R_2012,GR_mu,GM_ent}
  for some applications of deformed oscillators.}
 defined as $\Phi(N)\equiv a^\dagger a$, and the basic
 commutation relation for the generating elements $a$, $a^\dagger$
 and $N$ are of the form
\[ \Phi(N)=\frac{N}{1+\tilde{\mu} N} \ ,
\]
\begin{equation}                                  \label{mu-commut}
a a^\dagger - a^\dagger a  = \Phi(N+1) - \Phi(N) =
\frac{N+1}{1+\tilde{\mu} (N+1)} - \frac{N}{1+\tilde{\mu} N} \ .
\end{equation}
 The related position and momentum operators              taken in\,\cite{Jan}
 as $X=a+a^\dagger$ and $P={\rm i} (-a+a^\dagger)$ commute according to the
 formula
\begin{equation}                                         \label{mu-H}
[X, P] = {\rm i} \hbar \frac{2(1 - \tilde{\mu} {\cal
H}/\hbar\omega)^2}{1+\sqrt{1+ \tilde{\mu}^2(1-\tilde{\mu} {\cal
H}/\hbar\omega)^2}}\
\end{equation}
 which supplies a particular nontrivial realization of eq.~(\ref{f-HA}).
The relation (\ref{mu-H}), with the Hamiltonian expressed in terms
of $P$ and $X$, would lead to the corresponding (rather complicated)
modified version of uncertainty relation.  %which we do not present.

More recently, due to well known works of Kempf
                                          (see in particular\,\cite{Kempf})
and some others, the interest and attention to diverse
generalizations of the uncertainty relation directly connected with
modified HAs, as well as to diverse new modifications of the HA have
 grown significantly,  mainly   under         influence\,\cite{Garay}
 of string theory and of (loop) quantum gravity. In Ref.\,\cite{Kempf}
  Kempf, starting with {\em multi-mode} system of deformed
  oscillators                                covariant\,\cite{P-W}
  with respect to the quantum group $U_q(n)$, and given
  by the relations of commutation for generating elements,
  \[
\begin{array}{l}
 a_k a_j - {q}^{-1/2} a_j a_k = 0, \ \ \ k>j , \\
  a^\dagger_k a^\dagger_j - {q}^{1/2} a^\dagger_j a^\dagger_k = 0, \ \ \ k>j
 ,\\
  a_k a^\dagger_j - {q}^{1/2} a^\dagger_j a_k = 0, \ \ \ k\neq j
 ,\\
  a_k a^\dagger_k - q a^\dagger_k a_k = 1+(q -1) \sum_{j<k} a^\dagger_j a_j \
  ,  \end{array}
\]
 derived the associated {\em multidimensional} deformed Heisenberg algebra
 of $X_i$ and $P_i$, as well as the related generalized uncertainty relations.

 For our goals in this paper it  suffices to quote the 1-dimensional (or one-mode)
 form of the commutator,
\begin{equation}                                      \label{X2-P2}
[X, P] = {\rm i} \hbar +
   {\rm i} \hbar \left(q-1\right)\!\left(\frac{X^2}{4L^2}
   +\frac{P^2}{4K^2}\right) ,
\end{equation}
and the inferred respective modified uncertainty relation
\begin{equation}                                                \label{dxdp}
\Delta x \Delta p \geq \hbar [1+\beta (\Delta p)^2 + \gamma ]
\end{equation}
                                              (see Ref.\,\cite{Kempf}
 for the meaning of $L, K, \beta, \gamma$\ too).
Setting $L=K={1}/{\sqrt2}$ in eq.(\ref{X2-P2}) gives the Hamiltonian
of harmonic oscillator, and we come to the formula (\ref{mu-HA}).
 Besides, the relations (\ref{X2-P2})-(\ref{dxdp})
 can be deduced from the $q$-deformed commutation rule known from Ref.\,\cite{Arik},
  namely
\begin{equation}                     \label{AC}
a a^\dagger -q a^\dagger a = 1 \ .
\end{equation}
 There is an important immediate implication of the
 modified  uncertainty relation (\ref{dxdp}):
 the appearance of minimal uncertainty of the coordinate
 (minimal length) which is exactly expressed through the $\beta$.
 This fact leads to important physical consequence\,\cite{Garay,Hossen}).

Whereas the formula (\ref{mu-HA}) gives simplest possible nontrivial
realization (linear in ${\cal H}$) of the relation (\ref{f-HA}), on
the r.h.s. of eq.(\ref{mu-H}) there appears the expression which is
essentially nonlinear in the Hamiltonian ${\cal H}$.
   As a consequence, there should result a very
nontrivial uncertainty relation, obtained through fixation of
particular choice of the Hamiltonian ${\cal H}$ in terms of the
position and momentum operators.

In what follows we shall consider generalized HAs with more involved
(even in the single-mode case at hands) and relate them with
 nonlinear models of deformed oscillators.

%\vspace{1mm}
\section{$q$-Deformed Heisenberg algebra}

An alternative approach to deforming the Heisenberg algebra was
considered in the literature.
  Namely,                                              in\,\cite{Ch-K}
  and elsewhere, yet another class of deformed HA
  was studied such that deformation is introduced
  in the commutator in l.h.s. of the basic DR
  (this is the $q$-deformed HA):
\begin{equation}                                \label{Kl}
X P-q PX = {\rm i}\hbar \ .
 \end{equation}
  Below in this section we mainly follow the           work\,\cite{Ch-K}.
 The latter equality is required to be connected with
 specially deformed oscillator algebra with generating
 elements $a^+$, $a^-$   (creation/annihilation operators,
 not necessarily strict conjugates of each other) and $N$, such that
\begin{equation}                                        \label{N-a}
    [N,a^+]=a^+  \ , \hspace{15mm}  [N,a^-]=-a^- \ ,
\end{equation}
\begin{equation}                                         \label{H-G}
H(N)a^- a^+ - G(N)a^+ a^- = 1 \ .
\end{equation}
 The position and momentum operators are written as
\begin{equation}                                         \label{f-g}
X \equiv f(N) a^- + g(N) a^+ , %\hspace{4mm}
\end{equation}
\begin{equation}
P \equiv                                              \label{k-h}
 {\rm i}\left(
k(N) a^+ - h(N) a^-\right) .
\end{equation}
 For convenience, in our further treatment we put $\hbar = 1$.

On the base of (\ref{N-a}), for a function
 ${\cal F}(N)$ we have
\begin{equation}                                        \label{F(N)}
{\cal F}(N) a^{\pm} = a^{\pm} {\cal F}(N\pm 1), \hspace{7mm}
 [{\cal F}(N),a^\pm a^\mp] =0.
\end{equation}
Proceeding like                                     in~\cite{Ch-K}
and using (\ref{H-G})-(\ref{F(N)}), we deduce
\begin{equation}                                          \label{h/h}
\frac{ h(N\!+\!1)}{ h(N)} = q\, \frac{ f(N\!+\!1)}{ f(N)}\, ,
  \hspace{5mm}
  \frac{ k(N\!-\!1)}{ k(N)} = {q}\, \frac{ g(N\!-\!1)}{ g(N)}
\end{equation}
as well as the expressions
\begin{equation}                                         \label{H(N)}
H(N) = f(N) k(N+1) + q~ h(N) g(N+1) \ ,
\end{equation}
\begin{equation}                                       \label{G(N)}
G(N) =  g(N) h(N-1) + q~  k(N) f(N-1) \ .
\end{equation}
Solving (\ref{h/h}) yields
\begin{equation}                                        \label{f,k(N)}
f(N)\!=\! k(N)\!=\!\frac{1}{\sqrt 2}\, {q}^{N}\!, \ \
 h(N)\!=\! g(N)\!=\! \frac{1}{\sqrt 2} \, {q}^{2N} \ ,
\end{equation}
that leads to the relations
\begin{equation}                                       \label{H'(N)}
H(N) = \frac12\, {q}^{2N+1}\Bigl(1
 + q^{2N+2} \Bigr)\, ,
\end{equation}
%\end{document}
\begin{equation}                                        \label{G'(N)}
G(N) = \frac12\, {q}^{2N}\Bigl(1 + {q}^{2N-2} \Bigr)\, .
\end{equation}
With account of (\ref{f,k(N)}) put in (\ref{f-g})-(\ref{k-h}),
 $X$ and $P$ take the form
\[
X\!=\!\frac{1}{\sqrt 2} \, \Bigl(q^{2N} a^+ +  q^N a^- \Bigr) ,
\hspace{2mm}
 P\!=\!\frac{\rm i}{\sqrt 2} \, \Bigl(q^N a^+\!-\!q^{2N} a^-\Bigr).
\]
 The structure function $\Phi(N)$,            see Ref. \cite{Melj,Bona},
 which determines the bilinears (cf.(\ref{F(N)})),
\[
a^+ a^- = \Phi(N)\, , \ \ \ \ \   a^- a^+ = \Phi(N+1)\, ,
\]
 and the commutation relation
 \begin{equation}                                \label{a-a+}
[a^-, a^+]=\Phi(N+1)-\Phi(N),
\end{equation}
also gives the action formulas in the analog of Fock space:
\begin{equation}                                          \label{a+n}
a^+ |n\rangle \!=\! \sqrt{\Phi(n\!+\!1)} |n\!+\!1\rangle ,
 \ \ \ a^-|n\rangle \!=\! \sqrt{\Phi(n)} |n\!-\!1\rangle\ .
\end{equation}
It can be derived from the functions $H(n)$ and $G(n)$ in
(\ref{H-G}) according to the recipe
\begin{equation}                                          \label{Phi-GH}
{\Phi(n)} = \frac{G(n-1)!}{H(n-1)!}\biggl(\frac{1}{H(0)} +
\sum^{n-1}_{j=1} \frac{H(j-1)!}{G(j)!} \biggr) \
\end{equation}
where $F(n)! \equiv\!F(n) F(n\!-\!1) \ldots F(2) F(1)$ and
$F(0)!=\!1$.
 As result, from (\ref{H'(N)}),(\ref{G'(N)}) and (\ref{Phi-GH})
 one obtains the SF which allows to cast (\ref{H-G})
 into the form  (\ref{a-a+}):
\begin{equation}                                           \label{Phi(n)}
{\Phi(n)} =  \frac{2 q^{-n}}{(1+q^{2n-2})
(1+q^{2n})}\left(1+\frac{q^{n}-q^{-n+1}}{q-1}\right) \, .
\end{equation}
Note that this SF essentially differs from that of the Arik-Coon
                                               $q$-oscillator \cite{Arik}
 whose structure function is $\varphi_{AC}(n)=(q^n-1)/(q-1)$
 and also from the Biedenharn-Macfarlane       $q$-oscillator \cite{Bied,Mcf}
 whose SF is $\varphi_{BM}(n)=(q^n-q^{-n})/(q-q^{-1})$.

 If $q\to 1$ we have $f=g=h=k=1/{\sqrt
2}$,\ $G=H=1$, \ the structure function $\Phi(n)=n$ of usual
oscillator and the known relations $X=\frac{a^+ + a^-}{\sqrt 2}, \
  P = \frac{{\rm i}(a^+ -a^-)}{\sqrt 2}$.

%\bigskip
 \section{A $q,p$-deformation of the Heisenberg algebra }
%  \medskip

As our further extension of (\ref{Kl}) let us consider the
two-parameter deformed version of HA of the form
\begin{equation}                                       \label{pqHA}
    p X P-q PX = {\rm i}\hbar \ .
\end{equation}
In analogy to (\ref{h/h}), (\ref{H(N)}) and (\ref{G(N)}) we obtain
formulas
%{\cal F}(N)
\begin{equation}                                        \label{f'/f'}
\frac{\tilde{f}(N\!+\!1)}{ \tilde{f}(N)} \!=\!
 \frac{p}{q} ~ \frac{\tilde{h}(N\!+\!1)}{ \tilde{h}(N)} \, ,
  \hspace{4mm}
\frac{ \tilde{g}(N\!-\!1)}{ \tilde{g}(N)} \!=\! \frac{p}{q} ~ \frac{
\tilde{k}(N\!-\!1)}{ \tilde{k}(N)} \, ,
\end{equation}
 and the expressions
\[ %begin{equation}
\tilde{H}(N) = p~ \tilde{f}(N) \tilde{k}(N+1) + q~\tilde{h}(N)
\tilde{g}(N+1) \ ,
\] %end{equation}
\[ %begin{equation}
\tilde{G}(N) = p~\tilde{g}(N) \tilde{h}(N-1) + q~ \tilde{k}(N)
\tilde{f}(N-1) \ .
\]
Setting $Q\equiv q/p$, as solution of (\ref{f'/f'}) we find
 \vspace{-2mm}
\[
\tilde{f}(N) \!=\! \tilde{k}(N) \!=\! \frac{1}{\sqrt 2} Q^{N}\!\!\!,
\ \ \ \ \tilde{h}(N) \!=\! \tilde{g}(N)\!=\! \frac{1}{\sqrt 2}
Q^{2N} \
\]
 \vspace{-2mm}
  that leads to the following result:
\begin{equation}                                           \label{H''(N)}
\tilde{H}(N) = \frac12 q \, Q^{2N}\Bigl( 1 + Q^{2N+2} \Bigr)\ ,
\end{equation}
\vspace{-2mm}
\begin{equation}                                           \label{G''(N)}
\tilde{G}(N) = \frac12 p\, Q^{2N}\Bigl( 1 + Q^{2N-2} \Bigr)\ .
\end{equation}
%\vspace{-2mm}
 Also, from (\ref{f-g}), (\ref{k-h}) we deduce $X$ and $P$ in the form (compare with
 the formulas just after eq. (\ref{G'(N)})):
\[
 \vspace{-1mm}
X = \frac{1}{\sqrt 2} \, \Bigl[Q^{2N}\!a^+
 +  Q^N\!a^- \Bigr]  ,  \hspace{4mm}
P = \frac{\rm i}{\sqrt 2} \,
 \Bigl[
Q^N\!a^+ -
  Q^{2N}\!a^-\Bigr] .
 \]
  The SF $\tilde\Phi(N)$ which enters the nonlinear
  relationship of $a^\pm a^\mp$ with $N$          similar to (\ref{a-a+})
and determines the corresponding action formulas for
 $a^+$,\ $a^-$ similar to (\ref{a+n}), can be derived from eq. (\ref{Phi-GH})
using the functions $\tilde{H}(n)$ and $\tilde{G}(n)$ in
(\ref{H''(N)})-(\ref{G''(N)}).
 As result, for $\tilde\Phi$ we obtain    %on the base of (24) and (25)
\[
{\tilde\Phi(n)} \!=\! \frac{2 p^{-1} Q^{-n}}{(1+Q^{2n-2})
(1+Q^{2n})}
 \left(1\!+\!\frac{Q^{n}\!-\!
  Q^{-n+1}}{Q-1}\right) =
  \]
\begin{equation}                                           \label{Phi'(n)}
 %{\tilde\Phi(n)}
 \ \ \ =\! \frac{2 q^{-n} p^{5n-3}}
                {(q^{2n-2}+p^{2n-2})(q^{2n}+p^{2n})}
 \left(1\!+\!\frac{[2n\!-\!1]_{q,p}}{(qp)^{n-1}}\right)
\end{equation}
where $[m]_{q,p}\equiv \frac{q^m-p^m}{q-p}$ denotes the $q,p$-number
corresponding to a number $m$.

Formula (\ref{Phi'(n)}) gives the SF of nonstandard two-parameter
deformed quantum oscillator. It is nonsymmetric under
$q\leftrightarrow p$ (because of the factor $q^{-n}p^{5n-3}$ in the
numerator) and thus obviously differs from the well known
                                           $q,p$-oscillator \cite{Chakra,Arik-2}
 whose structure function $\varphi_{q.p}(n)=[n]_{q,p}$ is
($q\leftrightarrow p$)-symmetric.

 With the Hamiltonian taken in the conventional form
\[  %begin{equation}
 {\cal H}=\frac12 (a a^+ + a^+ a)=
  \frac12 \Bigl({\tilde\Phi(N+1)}+{\tilde\Phi(N)}\Bigr)\, ,
\]  %end{equation}
the energy spectrum is $E(n)= \frac12
\bigl({\tilde\Phi(n+1)}+{\tilde\Phi(n)}\bigr)$.

Note that at $p\to 1$ the results obtained here
 for the $p,q$-deformed HA reduce to those of the
 preceding section (say, (\ref{Phi'(n)}) reduces to (\ref{Phi(n)})),
 whereas for the case $p=q\neq 1$  we come to  the
 structure function $\phi(n)\!=\!\frac{n}{q}$,
 with $X$ and $P$ the same as those given in the last
 line of Sec.3. Obviously, now again we deal with usual harmonic oscillator,
 but the spacing in its (linear) energy spectrum is $\frac1q$\,-\,scaled.

 \section{Two-sided (or three-parameter) deformed Heisenberg algebra}
 \medskip

And now let us study our main object -- two-sided, three-parameter
generalization of the HA in which the deformation (\ref{pqHA}) is
combined with the modification (\ref{mu-HA}) of the ordinary HA
  (\ref{HA}).
    That is, we will deal with the relation
\begin{equation}                                             \label{qp-H}
\breve{q} X P -\breve{p} P X =
{\rm i} \hbar (1 + \mu {\cal H})\
\end{equation}
and link it with a nonlinear oscillator algebra such that
\begin{equation}                                           \label{31}
\breve{H}(N)a^- a^+ - \breve{G}(N)a^+ a^- = 1 ,
\end{equation}
 and the relations (\ref{N-a}),(\ref{F(N)}) do hold.
  Like before, we put
\begin{equation}                                          \label{32}
X \!=\! \breve{f}(N) a^- \!+\! \breve{g}(N) a^+ , \hspace{3mm}
 P \!=\! {\rm i}\bigl( \breve{k}(N) a^+ \!-\! \breve{h}(N) a^-\bigr).
\end{equation}
 Set $\frac{\breve{q}}{\breve{p}}\!=\!\breve{Q}$.
   Redefining $ X\!\to\!\tilde{X}\!\equiv\!\sqrt{\breve{p}}\, X $, \
   $P\!\to\!\tilde{P}\!\equiv\!\sqrt{\breve{p}}\, P  $,
     it is seen that instead  of (\ref{qp-H}) we  have
             (compare with (\ref{mu-HA}) and (\ref{Kl})):
\begin{equation}                                           \label{33}
 \tilde{X} \tilde{P} - \breve{Q} \tilde{P} \tilde{X} =
  {\rm i} \hbar (1 + \mu {\cal H})\ .
\end{equation}
By analogy with preceding analysis we deduce
\begin{equation}                                             \label{34}
   \frac{ \breve{f}(N\!+\!1)}{\breve{f}(N)}\!=\! \breve{Q}^{-1}
  \frac{\breve{h}(N\!+\!1)}{\breve{h}(N)}  ,
\hspace{4mm}  \frac{\breve{g}(N\!-\!1)}{\breve{g}(N)}\!=\!
\breve{Q}^{-1} \frac{\breve{k}(N\!-\!1)}{\breve{k}(N)} \
\end{equation}
which is formally the same as (\ref{G''(N)}), and also
\begin{equation}                                             \label{35}
\breve{H}(N) = \breve{q}\,\breve{h}(N)\breve{g}(N+1) +
\breve{p}\,\breve{f}(N)\breve{k}(N+1) -\frac{\mu}{2} \ ,
\end{equation}
\begin{equation}                                             \label{36}
\breve{G}(N) =
\breve{q}\,\breve{k}(N)\breve{f}(N-1)+\breve{p}\,\breve{g}(N)
\breve{h}(N-1) +\frac{\mu}{2}\,.
\end{equation}
Like above, the solutions of (\ref{34}) are
\begin{equation}                                             \label{37}
 \breve{f}(N)\!=\!\breve{k}(N)\!=\!\frac{1}{\sqrt2}\, \breve{Q}^{N}\!, \ \ \ \
 \breve{h}(N)\!=\!\breve{g}(N)\!=\!\frac{1}{\sqrt2}\, \breve{Q}^{2N}  ,
\end{equation}
that leads to the result:
\begin{equation}                                              \label{38}
\breve{H}(N) = \frac{\breve{q}}{2}\, \breve{Q}^{2N}\left(1 +
\breve{Q}^{2N+2}\right)-\frac{\mu}{2}\ ,
\end{equation}
\begin{equation}                                             \label{39}
\breve{G}(N) = \frac{\breve{p}}{2} \breve{Q}^{2N}\left( 1 +
\breve{Q}^{2N-2}\right)+
 \frac{\mu}{2}\ .
\end{equation}
Substituting (\ref{37}) into (\ref{32}) yields the expressions for
the position and momentum operators whose form coincides with $X$
and $P$ given immediately after eq. (\ref{G''(N)}).

  From (\ref{Phi-GH}), with account of (\ref{38}), (\ref{39}) we have,
  in principle, the corresponding structure function $\breve{\Phi}(n)$.
  It depends on $N$, on the parameters $\breve{q}$,\ $\breve{p}$,\ $\mu$
 of deformation, and through the formula ${\cal H}=\frac12\breve{\Phi}(n+1) +
  \frac12\breve{\Phi}(n)$ yields the spectrum of Hamiltonian.
 However, the very summation in    (\ref{Phi-GH})
 for present case is in general problematic.
  For that reason, we will consider more tractable specified cases.

  Consider special situation:
  $\breve{H}(N)=\breve{G}(N)$, \ $\breve{q}\neq\breve{p}$.
   Then, from (\ref{38})-(\ref{39}) we have
 \[ %begin{equation}
\mu=\frac12 \breve{p}\,\breve{Q}^{2N}
\left[\breve{Q}-1+\breve{Q}^{2N-2}(\breve{Q}^5-1) \right],
\]  %end{equation}
\[
\breve{H}(N)=\breve{G}(N)=\frac14 \breve{p}\,\breve{Q}^{2N} \left[
\breve{Q}+1+\breve{Q}^{2N-2}(\breve{Q}^5+1) \right].
\]
Putting this in (\ref{Phi-GH}) and performing necessary
calculations, we are able to obtain the explicit form of SF in this
special case of $\breve{H}=\breve{G}$, \ $\breve{q}\neq\breve{p}$,
namely:
\[\breve{\Phi}(n)=\frac{4\breve{Q}^2}{\breve{p}(1+\breve{Q}^2)(1+\breve{Q}^3)}
- \frac{4}{\breve{p}(1+\breve{Q})}
\biggl(\frac{1-\breve{Q}^{2-2n}}{1-\breve{Q}^2}+\biggr.
\]
\[
\biggl. +
\sum^{n-1}_{j=1}\frac{1+\breve{Q}^5}{\breve{Q}^2(1+\breve{Q})+\breve{Q}^{2j}(1+\breve{Q}^5)}
\biggr).
\]
The obtained formula gives the SF of nonstandard two-parameter
($\breve{q},\breve{p}$-)deformation of quantum harmonic oscillator.
If in addition $\breve{Q}=1$ \ (i.e. $\breve{q}=\breve{p}$), then
 $\mu=0$, $\breve{H}=\breve{G}=\breve{q}$, and we have
 $\breve{\Phi}(n)=\frac{n}{\breve{q}}$ (compare with the last paragraph of Sec.4).

\section{Unusual $N$-dependence of the factor $\mu$ (and $q$) }

 We are interested in finding direct link of the just explored
 $\breve{q},\breve{p},\mu$-deformed HA,               see (\ref{qp-H}),
 with some better known, well-studied version of deformed
 oscillator algebra.
   Namely, let us require that the results of the preceding
 Section would reproduce the well-known              model \cite{Chakra,Arik-2}
 of $q,p$-oscillator\footnote{For some
 applications of $q,p$-oscillators see e.g. \cite{AdGa,SIGMA,G-R_2012}}
 whose main defining relations are
\begin{equation}                                            \label{40}
a^- a^+  - q a^+ a^- = p^N
\end{equation}
along with analog of       (\ref{N-a}).   In view of   (\ref{31}),
 that means that the following equalities should hold:
\begin{equation}                                          \label{41}
 \breve{H}(N) = p^{-N}  ,   \hspace{10mm}
 \breve{G}(N) = q p^{-N} \ .
\end{equation}
 Validity of the latter will guarantee that now the SF
 ($q,p$-symmetric) will coincide with
 $\phi_{q,p}(n)\equiv[n]_{q,p}=\frac{q^n-p^n}{q-p}$ from \cite{Chakra,Arik-2}.

 Equating (\ref{41}) to (\ref{38})-(\ref{39}) for $\breve{H}(N)$
 and $\breve{G}(N)$ leads us to the desired relations
\begin{equation}                                             \label{42}
{\mu} = {\mu} (\breve{q},\breve{p},q,p;N) = \breve{q}\,
 \breve{Q}^{2N}\left(1+\breve{Q}^{2N+2}\right)-2p^{-N}  ,
\end{equation}
\begin{equation}                                             \label{43}
{\mu} \!=\! {\mu} (\breve{q},\breve{p},q,p;N) \!=\! 2 {q}{p}^{-N}
-\breve{p}\, \breve{Q}^{2N}\left(1 \!+\! \breve{Q}^{2N-2} \right),
% Q^{-4N+3} + Q^{-2N+1}-2 q p^{-N}\ ,
\end{equation}
from which, for $q\neq -1$, we deduce the formula
\begin{equation} {\mu}(\breve{q},\breve{p},q;N) =               \label{44}
 \frac{\breve{p}\, \breve{Q}^{2N}\!\left[
\breve{Q}^{2N-2}(q\breve{Q}^5 -1)+q\breve{Q}-1\right]}{1+q}\ .
\end{equation}
From (\ref{42})-(\ref{43}) we also obtain the relation which
expresses the parameter $q$ in (\ref{40}) through $p$ and the
deformation parameters $\breve{p}, \breve{Q}$ (without $\mu$):
\begin{equation}                                               \label{45}
{q}\!= -1\!+ \frac{\breve{p}}{2}
 p^{N}\breve{Q}^{2N}\!\left[1+\breve{Q}+\breve{Q}^{2N-2}(1\!+\!\breve{Q}^5) \right].
\end{equation}
Here we come to the basic point: though in the starting relation
(\ref{qp-H}) the parameter $\mu$ was understood as constant, the
result of linking with $q,p$-oscillator (\ref{40}) inevitably leads
to appearance of the obvious dependence of $\mu$, besides the
parameters $q$, $p$ and $\breve{q}$, $\breve{p}$, also on the number
operator $N$, see (\ref{42})-(\ref{44}). The same can be said about
the deformation "parameter"\ $q$ in (\ref{45}), whereas  $p$ can
still be viewed as true, constant parameter.
 Note that yet another formula for the parameter $q$ follows also
 directly from (\ref{43}):
\begin{equation}                                          \label{46}
{q}\!= \frac{1}{2}\,p^{N}\!\left[\,\mu+\breve{p}\,
\breve{Q}^{2N}\left(1 \!+\! \breve{Q}^{2N-2} \right)  \right] .
\end{equation}

Let us emphasize the following. Formulas (\ref{44})-(\ref{46})
relate deformation parameters that enter different relations:
(\ref{qp-H}) on one hand and (\ref{40}) on the other hand. That is,
the two-sided deformed HA can be presented by the $q,p$-oscillator
algebra (\ref{40}), but then we have to put in (\ref{40}), besides
the usual parameter $p$, the $N$-dependent parameter $q=q(N)$ given
by (\ref{45}) or (\ref{46}). The latter formulas assume also an
interesting possibility: we can express the deformation parameter in
(\ref{40}) in terms of the 3 parameters of deformed HA (\ref{qp-H}).
Namely, we infer the formulas:
 \begin{equation}                                              \label{47}
q=\frac{\breve{p}\,\breve{Q}^{2N}\bigl(1+\breve{Q}^{2N-2}\bigr)+\mu}
       {\breve{p}\,\breve{Q}^{2N+1}\bigl(1+\breve{Q}^{2N+2}\bigr)-\mu} ,   \hspace{8mm}
%%\]
%and also
%%\[
  p^N=2\left[\breve{p}\,\breve{Q}^{2N+1}\bigl(1+\breve{Q}^{2N+2}\bigr)-\mu\right]^{-1}.
 \end{equation}
In summary, we conclude that in the formulas (\ref{qp-H}),
(\ref{33}), (\ref{35})-(\ref{36}) and (\ref{38})-(\ref{39}) there
unavoidably appears novel feature - dependence of $\mu$ on $N$.
Likewise, the parameter $q$ in (\ref{40}) also becomes depending on
$N$, as follows from (\ref{45}), (\ref{46}) or (\ref{47}).
 These unusual features are direct consequence of presenting
 (\ref{qp-H})-(\ref{31}) through (\ref{40}),
 or of imposing $\breve{H}(N)=\breve{G}(N)$, see above.

\newpage
\underline{\it Some reduced cases}.

\begin{itemize}
  \item       If $\breve{Q}\to 1$, formulas (\ref{44})-(\ref{45}) give  %reduce to
 $\mu ={2\breve{q}(q-1)}/{(q+1)}$ and $q=-1+2\breve{q}p^{N}$, from which
$\mu=2(\breve{q}-p^{-N})$. Then, sending $q\to 1$ (equivalently
$p^{-N}\to\breve{q}$) we arrive at $\mu = 0$.
 That is, $\breve{Q}\to 1$ and $q\to 1$ agree with $\mu\to 0$.
Recall that $\breve{Q}$ (i.e. $\breve{q}/\breve{p}$) resp. $q$ enter
different relations: eq.(\ref{qp-H}) resp. eq.(\ref{40}).
   \item   We can link two parameters, one of (\ref{qp-H}) and one of (\ref{40}),
 by setting $q=\breve{Q}^{-1}$.
  Then, the parameter $\mu$ takes simpler form:
   $\mu=\breve{q}\,(\breve{Q}-1)(\breve{Q}^2+1)\breve{Q}^{4N-2}$.
 Besides emphasized $N$-dependence, it depends on $\breve{q}$ and
 $\breve{p}$ (or $\breve{q}$ and $\breve{Q}$) from (\ref{qp-H}) only.
     \item     Let us examine the situation when we wish to realize the
  two-sided deformed HA not by  $q,p$-deformed oscillator, but
  by known one-parameter $q$-oscillator.
  Say, for Arik-Coon type $q$-oscillator, $a^-a^+ -qa^+a^-=1$ and the SF is
  $\varphi_{AC}(n)=
  \frac{q^n-1}{q-1}$.
  In this case, $H=1$, $G=q$ (see (\ref{41})), and the formulas (\ref{42})-(\ref{46}) slightly
  modify ($p^N$ and $p^{-N}$ disappear). Say, for $\mu$ we have
  $\mu=-2+\breve{p}\breve{Q}^{2N+1}(1+\breve{Q}^{2N+2})$.
\end{itemize}

\section{Conclusions}

 Our main results are contained in Sects.4-6 and give presenting of
 the $q,p$-deformed (one-sided) HA and similarly the $(q,p;\mu)$-deformed
 two-sided modification of HA in terms of respective
 {\it non-standard} deformed oscillators determined by their
 respective structure functions $\tilde\Phi(n)$ and $\breve\Phi(n)$.
 Note that the aspects concerning conjugation of $a^-$ and $a^+$,
 (non-)Hermiticity of $X$, $P$ and ${\cal H}$ are rather non-trivial
 in our setting and require special analysis (say, in analogy with \cite{Bagchi}).
  There is also an interesting question concerning modified versions
  of the Heisenberg uncertainty relation for the above studied
  deformations of HA. These issues will be studied elsewhere.

 We have also shown it is possible to link directly the {\it non-standard}
  deformed oscillator with the structure function
  $\breve\Phi(n)$, given by (\ref{Phi-GH}) plus (\ref{38})-(\ref{39}),
  with the known $q,p$-deformed one.
  The price of that is:
  we come to the property that there appears {\em unavoidably}
  the dependence of $\mu$ (and $q$) on $N$.
 We may ask if such a dependence, due to deformation, of
 $\mu$ (and $q$) on $N$ is something really unusual, without precedents?
 The answer is "no" since more or less similar situations
 already appeared in deformed systems theory.
 Let us mention some of them.

1) From the early papers, e.g.                  \cite{Bied,Mar-Del}
and also from more recent ones as say \cite{GR-1},
 we know that in generalized uncertainty relation connected with a
 deformed quantum oscillator, there appears minimum uncertainty necessarily
 depending on the quantum number $n$, the eigenvalue of the operator
 $N$ on Fock basis state $|n\rangle$.

2) Next, let us mention that for such known deformed versions
 of quantum oscillator as BM, TD,
  and $q,p$-deformed             models \cite{Bied,Mcf,Odaka,Chatur,Chakra,Arik-2},
 the r.h.s. of the $a^-, a^+$ commutation relation contains
 the factors $q^{-N}$, $q^N$ or $p^N$, which turn into
 the constant "1" \ as $q\to 1$ or $p\to 1$.

  3) For the nonstandard $q$-analog of
the Lie algebra $so(3)$, the well-known constant 'classical'\
coefficient $\frac12$ in their representation formulas deforms
  into the special                             \cite{Gavr-93,Gavr-Kl}
   $m$-dependent multiplier
 $\left(\frac{[m]_q[m+1]_q}{[2m]_q[2m+2]_q}\right)^{1/2}$
 (when $q\to 1$, we recover $\frac12$).
  The same feature characterizes the formulas for representations \cite{Gavr-Io}
of nonstandard $q$-deformed analogs of higher rank Lie algebras
$so(n)$, \ $n>3$.

4) Finally there is some analogy with the appearance of novel
  $n$-dependence in recursion relations characteristic for
  "quasi-Fibonacci oscillators",\                       see \cite{G-K-R}.
  Namely, whereas we have  three-term recursion
relation with {\it  constant coefficients} for the energies of
  Fibonacci oscillators                                \cite{Arik-2}
  such as the already mentioned AC, BM, TD, and
  $q,p$-oscillators                             \cite {Arik,Bied,Mcf,Odaka,Chatur,Chakra},
 in the case of "quasi-Fibonacci oscillators"\ the coefficients
 in recursion relations for their energies
 necessarily \cite{G-K-R} depend on $n$, the eigenvalue of operator $N$
 on the state $|n\rangle$.

%\vspace{1mm}
\section*{Acknowledgement}

This work was partially supported by the Special Programme of
Division of Physics and Astronomy of NAS of Ukraine.

\end{document}